\documentclass[a4paper,fleqn,usenatbib]{mnras}

\usepackage{newtxtext,newtxmath}

\usepackage[T1]{fontenc}
\usepackage{ae,aecompl}

\usepackage{graphicx}	
\usepackage{amsmath}	
\usepackage{amssymb}	
\usepackage{float}

\def\hi{H\,{\sc i}}
\def\deg{$^{\circ}$}
\def\kms{km~s$^{-1}$}
\def\msun{$M_{\odot}$}

\def\Mb{$M_\mathrm{b}$}
\def\jb{$j_\mathrm{b}$}
\def\Jb{$J_\mathrm{b}$}
\def\jm{$j_\mathrm{b}-M_\mathrm{b}$}

\title[The \jm\ relation in low-mass galaxies]{The relation between specific baryon angular momentum and mass for a sample of nearby low-mass galaxies with resolved \hi\ kinematics}

\author[E. C. Elson]{
E. C. Elson,$^{1}$\thanks{E-mail: elson.e.c@gmail.com}
\\
$^{1}$Department of Astronomy, University of Cape Town, Private Bag X3,
Rondebosch 7701, South Africa.
}

\pubyear{2015}

\begin{document}
\label{firstpage}
\pagerange{\pageref{firstpage}--\pageref{lastpage}}
\maketitle

\begin{abstract}
This paper investigates the relationship between  specific baryon  angular momentum \jb\ and baryon mass \Mb\ for a sample of nearby late-type galaxies with resolved \hi\ kinematics. This work roughly doubles the number of galaxies with $M_\mathrm{b}\lesssim 10^{10}$~\msun\ used to study the \jm\  relation.  Most of the galaxies in the sample have their baryon mass dominated by their gas content, thereby offering \jb\ and \Mb\ measures that are relatively unaffected by uncertainties arising from the stellar mass-to-light ratio.  Measured \hi\ surface density radial profiles together with optical and rotation curve data from the literature are used to derive a best-fit relation  given by $j_\mathrm{b}=qM_\mathrm{b}^{\alpha}$, with $\alpha=0.62\pm 0.02$ and $\log_{10}q=-3.35\pm 0.25$.  This result is consistent with the $j_\mathrm{b}\propto M_\mathrm{b}^{2/3}$ relation that is theoretically expected and also measured by \citet{OG14} for their full sample of THINGS spiral galaxies, \textcolor{black}{yet differs to their steeper relation found for subsets with fixed bulge fraction}.  The 30~arcsec spatial resolution of the \hi\ imaging used in this study is significantly lower than that of the THINGS imaging used by \citet{OG14},  \textcolor{black}{yet the results presented in this work are clearly shown to contain no significant systematic errors due to the low-resolution imaging.} 
\end{abstract}


\section{Introduction}
Important insights into the structural properties, formation histories and dark matter halo properties of galaxies can be gleaned from their angular momentum properties.  In the hierarchical galaxy formation scenario, angular momentum is thought to be transferred to collapsing proto-galaxies by the tidal field associated with an irregular matter distribution \citep{Peeble1969}.  

For stellar disks,   \citet{Takase1967} and \citet{Freeman1970} were the first to study the link between total angular momentum $J_*$ and mass $M_*$, finding a power law relation $J_*\propto M_*^{7/4}$.  \citet{Fall1983} was the first to study the distribution of galaxies in the $j_*-M_*$ plane, where $j_*\equiv J_*/M_*$ is the stellar specific  angular momentum.  He showed a sample of spirals to follow the power-law relation $j_*\propto M_*^{\alpha}$, with $\alpha\approx 0.6$.  He also showed a sample of ellipticals to follow a roughly parallel trend in the $\log_{10}j_* - \log_{10}M_*$ plane, yet one which is roughly a factor of 6 lower.  \citet{Romanowsky_Fall2012} furthered the work of \citet{Fall1983} by estimating $j_*$ from kinematic and photometric data that extended to large radii and $M_*$ from near-infrared luminosities.  They found the ellipticals and spirals to form two parallel $j_*-M_*$ tracks, with log-slopes of $\sim 0.6$.  \citet{Fall_Romanowsky2013} re-visited the work of \citet{Romanowsky_Fall2012} by considering the effects of variable near-infrared stellar mass-to-light ratios correlated with $B-V$ colours.  They found both disk-dominated and elliptical galaxies to have $\alpha=0.6\pm 0.1$, but the offset between sequences increased by about 30~per~cent.

The studies of \citet{Fall1983, Romanowsky_Fall2012, Fall_Romanowsky2013} estimated $j_*$ from velocity width measurements together with an assumption about the shape of the rotation curve.  \citet{OG14} were the first to measure $J$ by integrating $dJ$ over spatially and kinematically resolved observations of galaxies.  Furthermore, \citet{OG14} were the first to consider the contribution of gas (atomic and molecular) to $J$, in addition to the stellar contribution.  For a sample of 16 spiral galaxies from THINGS, they demonstrated a strong correlation between baryon mass \Mb,  specific baryon angular momentum \jb, and  bulge mass fraction $\beta$.  In the \jm\ plane, their data are consistent with the range of theoretically predicted (\jb, \Mb) values for ``isolated spiral galaxies evolved without major mergers, abnormal feedback, or otherwise exotic histories''.  Their power-law relation $j_\mathrm{b}\propto M_\mathrm{b}^{\alpha}$, with $\alpha \approx 2/3$,  therefore also reproduces the exponent found by \citet{Romanowsky_Fall2012} who considered only the stellar components of \jb\ and \Mb.  \textcolor{black}{However, for a fixed bulge fraction, \citet{OG14} find the residual scaling to have a power-law index $\alpha\approx 1$.  This, they say,`` implies that more massive spiral galaxies tend to have higher bulge fractions than less massive ones''. }

All of the above-mentioned studies focussed on relatively high-mass galaxies.  Until very recently, the $j-M$ relation for dwarf galaxies has remained unexplored.  \citet{Butler_2017} present measurements of \jb\ and \Mb\ for a sample of 14 rotating dwarf Irregular galaxies from the LITTLE THINGS survey \citep{LittleTHINGS}.  In gas-dominated systems with $M_\mathrm{b}\lesssim 10^{10}$~\msun, \hi\ kinematics are  required to measure the amount of angular momentum contained in the outer galaxy.  \citet{Butler_2017} find their dwarf galaxies to have systematically higher \jb\ values than expected from the $j_\mathrm{b}\propto M_\mathrm{b}^{2/3}$ scaling relation for spirals.  They attribute this to  a significantly steeper $j-M$ relation for \hi, due to the systematic variation of \hi\ fraction with \Mb.  Most recently, \citet{CC17} measured \jb\ and \Mb\ for five gas-rich dwarf galaxies.  They compare the \jb~-~\Mb\ distribution of their galaxies to the \jb~-~\Mb\ relation for the low-bulge-fraction ($\beta<0.05$) subset of \citet{OG14}, and find the dwarfs to have significantly higher \jb.  This, they say, suggests a difference in the evolution of angular momentum in the smaller galaxies, compared to the more massive spirals.  

This works aims to further study the relation between \jb\ and \Mb\ in a sample of 37 late-type and dwarf galaxies from the WHISP survey with spatially and kinematically resolved \hi\ kinematics, thereby doubling the number of galaxies with $M_\mathrm{b}\lesssim 10^{10}$~\msun\ studied in this context.  Section~\ref{data} presents the data sets used to calculate \jb\ and \Mb, while Section~\ref{analysis} details the specific methods of measuring the quantities.  The results are presented together with a discussion in Section~\ref{results}.  Section~\ref{systematics} presents the details of a test carried out using THINGS imaging to quantify the significance of systematic errors in the \jb\ and \Mb\ measures presented in this work.  The conclusions of this study are given in Section~\ref{conclusions}.

\section{Data}\label{data}
The primary data set for this study comes from \citet{Swaters2009} who present rotation curves derived from \hi\ observations for a sample of 62 dwarf galaxies observed as part of the Westerbork \hi\ survey of Spiral and Irregular Galaxies (WHISP) project.   From the full WHISP sample, \citet{Swaters2009} define as dwarfs all galaxies with Hubble type later than Sd, as well as spiral galaxies of earlier Hubble type with an absolute $B$-band magnitude fainter than -17.  The typical spatial resolution of their \hi\ observations is $12''\times 12''/\sin \delta$, where $\delta$ is the declination of the target.  However, because the signal-to-noise ratios for most of their galaxies are too low to derive reliable rotation curves using standard methods, \citet{Swaters2009} use 30~arcsec resolution versions of their data to produce rotation curves.  The reader is referred to \citet{SwatersPaper1} for the full details of the procedure used to produce the \hi\ data cubes.

\citet{Swaters2009} adopt a two-step procedure to generate an initial estimate of each galaxy rotation curve.  First, they attempt fitting a tilted ring model \citep{Rogstad1974} to the velocity field of the galaxy.  When this is not feasible, they use an interactive procedure to iteratively fit the rotation curve and orientation parameters to a set of six position-velocity diagrams from the observed data cubes.  They then use this information to produce a three-dimensional model data cube which they compare directly to the data.   The input rotation curve is interactively adjusted until a satisfactory comparison to the data is obtained.  \citet{Swaters2009}  determine the correction for asymmetric drift to be smaller than 3~\kms\ at all radii for all but three galaxies in their sample.  Because the corrections are so small, they do not correct their measured rotation curves for asymmetric drift.  The reader is referred to \citet{Swaters2009} for the full details of their procedure used to generate the rotation curves.  \citet{Swaters2009} divide their derived rotation curves into four different categories of quality $q$.  A reliable rotation curve is indicated by $q=1$, whereas  $q=2$ and $q=3$ indicate uncertain and highly uncertain rotation curves, respectively.  Finally, $q=4$ indicates a case for which no rotation curve could be derived.   

Table~2 from \citet{Swaters2009} contains almost all of the data used in this work to calculate $j_b$ and \Mb\ for WHISP galaxies.  It presents their rotation velocities at one, two, three, and four disk scale lengths, as well as the rotation velocity at the last measured point and the radius of the last measured point, $R_\mathrm{last}$.  It also lists the $R$-band properties from \citet{SwatersPaper2}: absolute magnitude (corrected for Galactic foreground extinction, but not internal extinction), disk scale length, and central surface brightness.  It  presents the adopted distance for each galaxy (also discussed in \citealt{SwatersPaper2}).  Not provided in \citet{Swaters2009} are the  $B$-band absolute magnitudes of the  galaxies.  For this work, the $B$-band apparent magnitudes (corrected for extinction) are obtained from the Third Reference Catalog of Bright Galaxies, (RC3, \citealt{RC3}) and converted to absolute magnitudes using the distances listed in \citet{Swaters2009}.  The total \hi\ masses and \hi\ radii used in this work are taken from Table~A.2 of \citet{SwatersPaper1}.  The 30~arcsec \hi\ total intensity maps of the galaxies, available for download from \url{http://wow.astron.nl}, are used to derive an \hi\ mass surface density profile for each galaxy (see Section~\ref{HI_surface_density_profiles}). 

The final sample of 37 WHISP galaxies  used in this work consists of those galaxies in \citet{Swaters2009} with $q\le 2$, inclination $i\le75$\deg, a valid $B$-band apparent magnitude in RC3, and for which an \hi\ total intensity map could be downloaded.  

\section{Analysis}\label{analysis}
The optical and rotation curve data presented in Tables~2 and 3 of \citet{Swaters2009} together with the \hi\ radius measurements from \citet{SwatersPaper1}, the \hi\ surface density profiles measured (in this work) from the 30~arcsec \hi\ total intensity maps,  and the $B$-band magnitudes from RC3 constitute all the information required to generate measures of the specific baryon angular momentum \jb\ and total baryon mass \Mb\ for the galaxies.  For each galaxy, \Mb\ is calculated by numerically evaluating the following integral:
\begin{equation}
M_{\mathrm{b}}=\int_0^{R_{\mathrm{HI}}}\Sigma_\mathrm{b}(R)\cdot 2\pi R\cdot dR,
\label{eqn:Mb}
\end{equation}
where $R$ is galactocentric radius, $R_{\mathrm{HI}}$ is the radius at which the inclination-corrected \hi\ mass surface density reaches 1~\msun~pc$^{-2}$, and $\Sigma_b(R)$ is the baryon mass surface density in units of \msun~pc$^{-2}$.  In this work, only the stellar, \hi, and He contributions to the baryon mass are considered.  The  total baryon  angular momentum $J_\mathrm{b}$ perpendicular to the disk of each galaxy is calculated as:
\begin{equation}
J_b=\int_0^{R_{\mathrm{HI}}} 2\pi R\cdot \Sigma_\mathrm{b}(R)\cdot V_{\mathrm{PE}}(R)\cdot R\cdot dR,
\label{eqn:J}
\end{equation}
where $V_{\mathrm{PE}}(R)$ is a parameterisation of the observed \hi\ rotation curve.  The  specific baryon angular momentum is then calculated as $j_\mathrm{b}=J_\mathrm{b}/M_\mathrm{b}$.

The rest of this section presents the details of the procedures used to calculate $\Sigma_\mathrm{b}(R)$ and $V_\mathrm{PE}(R)$ for each galaxy. 

\subsection{Rotation curves}
The circular velocity profile of each galaxy is obtained by fitting the measured \hi\ rotation curve with the analytic function
\begin{equation}
V_{\mathrm{PE}} (R)= V_{\mathrm{0}}\left(1-e^{-R/R_{\mathrm{PE}}}\right)\left(1+{\alpha R\over R_{\mathrm{PE}}}\right).
\label{eqn:polyex}
\end{equation}
This is the Polyex model from \citet{Giovanelli_Haynes_2002}.  $V_{\mathrm{0}}$, $R_{\mathrm{PE}}$, and $\alpha$, respectively, determine the amplitude of the outer rotation curve, the exponential scale length of the inner rotation curve, and the slope of the outer rotation curve.  
\subsection{\hi\ mass surface density profiles}\label{HI_surface_density_profiles}
In this work, the total baryon mass surface density profiles are calculated as 
\begin{equation}
\Sigma_\mathrm{b}(R)=\Sigma_{\mathrm{HI+He}}(R)+\Sigma_*(R),
\label{eqn_Mb}
\end{equation}
where $\Sigma_\mathrm{HI+He}(R)$ is the measured \hi\ mass surface density profile scaled by a factor of 1.342 to account for the presence of Helium \citep{HI_He} and $\Sigma_*(R)$ is the measured stellar mass surface density radial profile.   $\Sigma_\mathrm{b}(R)$ has units of \msun~pc$^{-2}$. 

The measured \hi\ surface density profiles of almost all  the galaxies in \citet{Swaters2009} have an exponential behaviour in their outer parts.  \citet{SwatersPaper1} give the scale length of each galaxy as determined from a fit to its outer disk.  However, many of the galaxies exhibit a central depression of \hi, and \citet{SwatersPaper1} make no attempt to model this.  Simply extrapolating the fitted exponentials from the outer to the inner disks will result in  significant over-estimates of the central \hi\ masses.  Therefore, the \hi\ surface density profiles were re-measured using the 30~arcsec \hi\ total intensity maps.  The position angle and inclination of each galaxy was calculated by fitting an ellipse to a thin ribbon of \hi\ flux in the outer \hi\ disk of each galaxy.  These orientation parameters were then used to calculate the azimuthally-averaged \hi\ mass surface density in concentric ellipses of width 15~arcsec .  To properly calibrate the resulting profiles, they were each scaled by a constant factor to ensure that the total \hi\ mass within 3.2~$R$-band disk scale lengths was equal to $\pi \cdot \left(\left<\mathrm{\Sigma_{HI}}\right>_{3.2h}\right)^2$, where $\left<\mathrm{\Sigma_{HI}}\right>_{3.2h}$ is the average \hi\ mass surface density within 3.2~$R$-band disk scale lengths given in Table~A.2 of \citet{SwatersPaper1}.  In \citet{SwatersPaper1}, the shortest baselines for the WHISP observations are stated to be 36 or 72~m.  The observations are therefore less sensitive to structures more extended than 5 or 10~arcmin.  All but three of the galaxies used in this work are smaller than 5~arcmin, and all of them are smaller than 10~arcmin.  The total \hi\ fluxes are therefore generally expected to be well determined.

\subsection{Stellar mass surface density profiles}\label{stellar_surface_density_profiles}
The $R$-band properties of the galaxies given in Tables~2 and 3 of \citet{Swaters2009} are used to generate stellar surface brightness profiles.  These profiles represent only the exponential disk components of the galaxies, no attempt was made to model the central components.  However, relatively few of the galaxies used in this work show a significant central excess of light.  The radius lever arm in $\vec{r}\times\vec{v}$ is small near the centre of any  galaxy, meaning that and stellar mass contained within a bulge will not contribute significantly to the total specific baryon angular momentum.  Furthermore, most galaxies are \hi-dominated in terms of their baryon mass.  The  effects of neglecting the central light components are therefore expected to be negligible in terms of the  \jb\ and \Mb\ measures derived in this work.   The galaxy distances are used to convert the surface brightness profiles to units of L$_{\odot}$/pc$^{2}$.  In order to convert them to stellar mass surface density profiles (\msun/pc$^{2}$), an $R$-band mass-to-light (M/L) ratio is calculated for each galaxy using one of the prescriptions from \citet{Bell_deJong} relating optical M/L ratios and colours of integrated stellar populations.  The relation based on their mass-dependent formation model with bursts is used:
\begin{equation}
\log_{\mathrm{10}}\left({M/L\over M_{\odot}/L_{\odot}}\right)=0.851(B-R)-0.820.
\label{eqn:MtoL}
\end{equation} 

Finally, having generated $V_\mathrm{PE}(R)$ and $\Sigma_b(R)$ for each galaxy, $M_{\mathrm{b}}$ and $J_b$ are calculated using eqns.~\ref{eqn:Mb} and \ref{eqn:J}.

\section{Results and Discussion}\label{results}
\begin{figure}
	\includegraphics[width=1\columnwidth]{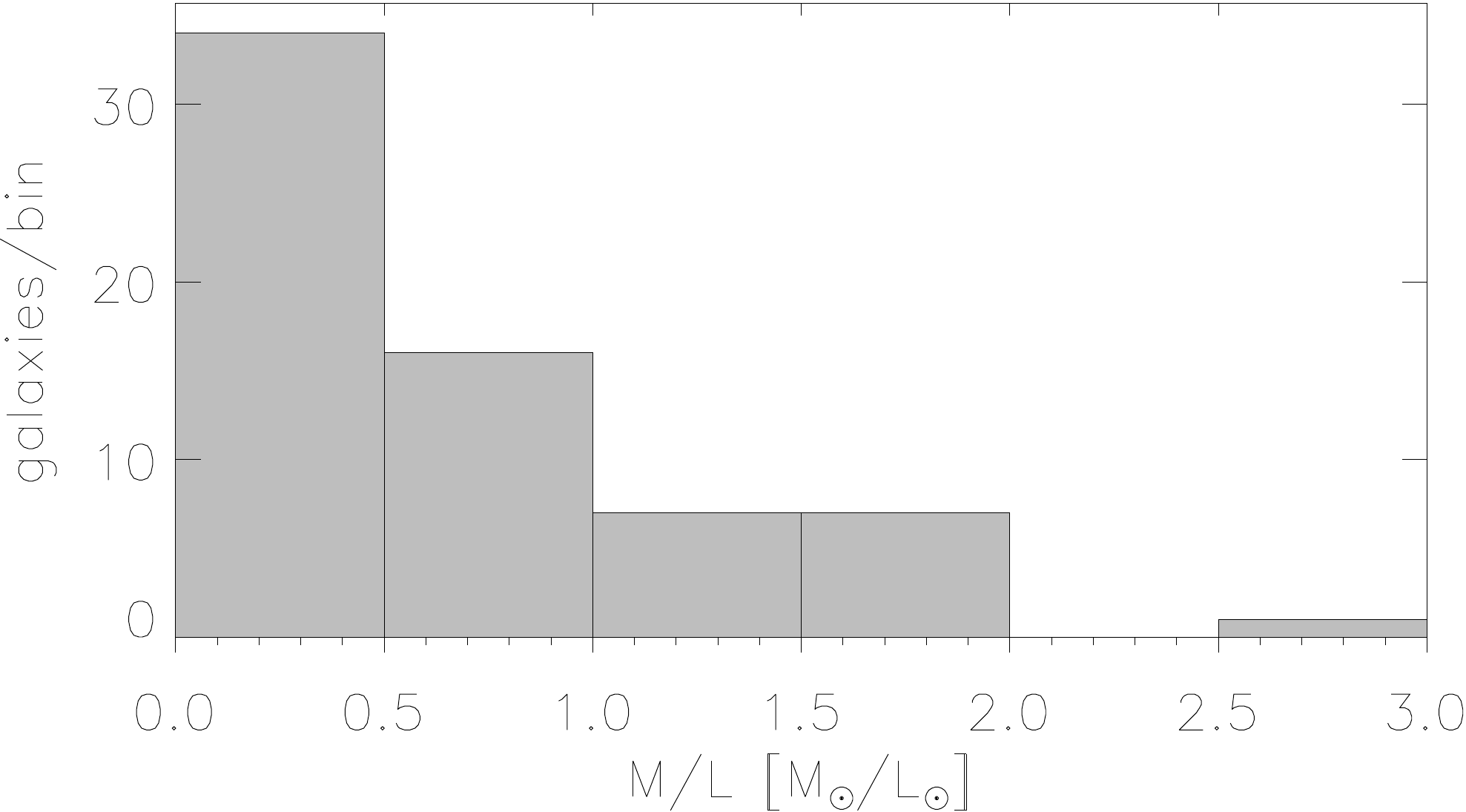}
    \caption{Distribution of $R$-band mass-to-light ratios for the galaxies used in this study, calculated using the prescription from \citet{Bell_deJong} given in Eqn~\ref{eqn:MtoL}.}
    \label{fig:M_over_L}
\end{figure}

\begin{figure}
	\includegraphics[width=1\columnwidth]{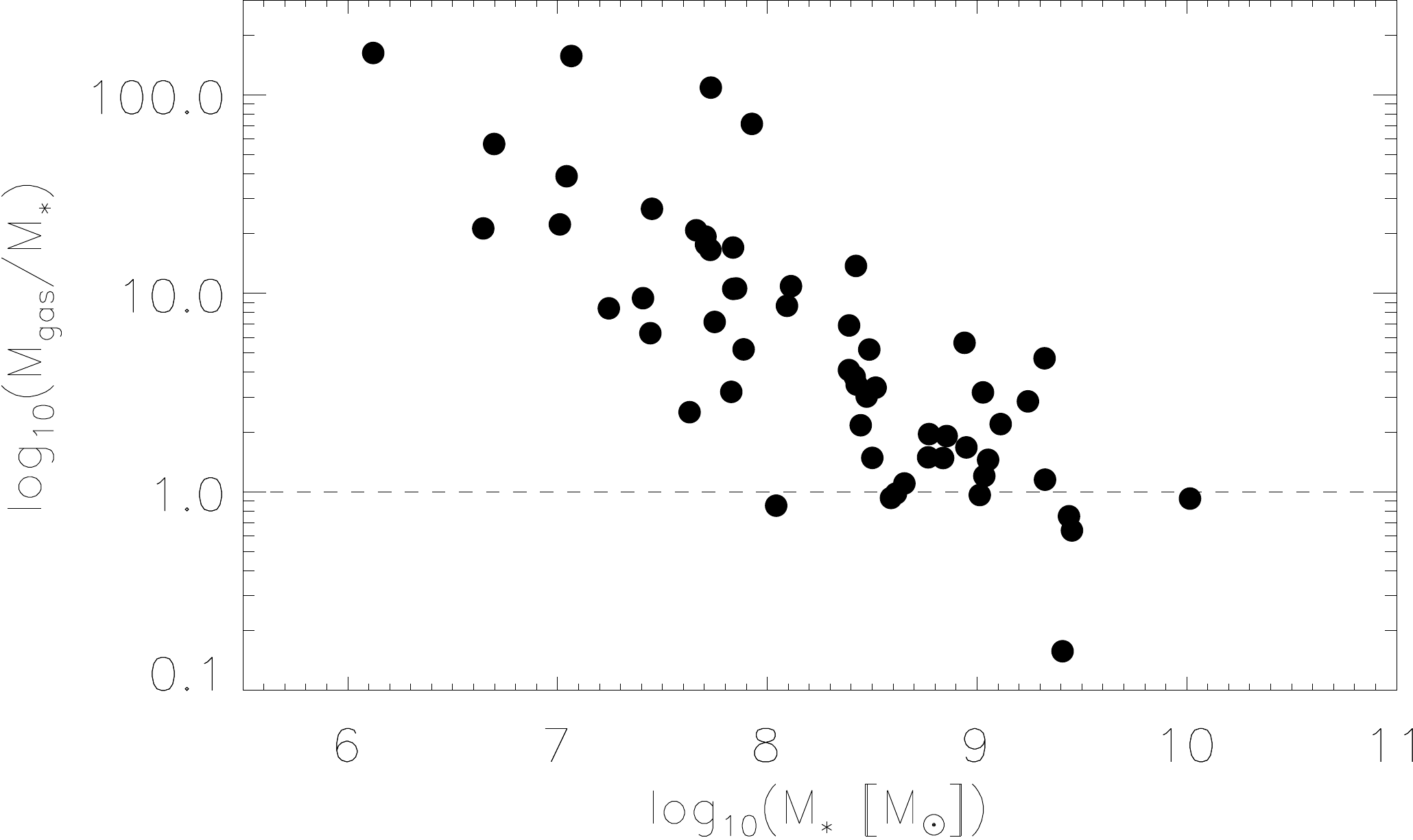}
    \caption{Gas fraction as a function of stellar mass for the galaxies used in this study.  $M_{\mathrm{gas}}=1.342\times M_{\mathrm{HI}}$ is used to account for the presence of helium and other metals.  Clearly, almost all galaxies with $M_*\lesssim 10^{9.3}$~\msun\ are gas dominated.}
    \label{fig:gas_frac}
\end{figure}

Figures~\ref{fig:profiles1} to \ref{fig:profiles4} show the $V_\mathrm{PE}(R)$, $\Sigma_\mathrm{HI+He}(R)$, $\Sigma_\mathrm{b}(R)$, and $j_\mathrm{b}(<R)$ profiles for all  galaxies used in this study.  Table~\ref{tab:results} lists the best-fit rotation curve parameters, $R$-band mass-to-light ratios, and \Mb, \Jb, \jb\ measures.  Some of the best-fit rotation curve parameters are clearly unphysical.  (e.g.~$R_\mathrm{PE}=1030$~kpc for UGC~1281), but they are used anyway given that main objective is to fit a smooth curve that accurately represents the measured rotation velocities.  The distribution of $R$-band M/L ratios is shown in Fig.~\ref{fig:M_over_L}.  Clearly, the large majority of galaxies have $\mathrm{M/L}<1~M_{\odot}/L_{\odot}$.  Fig.~\ref{fig:gas_frac} shows the gas fraction of each galaxy as a function of stellar mass.  Almost all galaxies with $M_*\lesssim 10^{9.3}$~\msun\ are gas dominated - many of them by a factor of 10 to 100.

\begin{table*}
	\centering
	\caption{Best-fit rotation curve parameters, mass-to-light ratios, and $M_b$, $J_b$ , $j_b$ measures for all the galaxies in this study.  Column~1 gives the  UGC number of each galaxy.  Columns 2 - 4 give, respectively, the amplitude (in units of \kms) of the outer rotation curve, the exponential scale of the inner rotation curve (in units of arcsec), and the slope of the outer rotation curve.   Column 5 gives the $R$-band mass-to-light ratio.  Columns 6 - 8 give, respectively, the total baryon mass (in units of 10$^8$~\msun), the  total baryon angular momentum (in units of 10$^{10}$~\msun~kpc~\kms), and the  specific baryon angular momentum (in units of kpc~\kms).}

	\label{tab:results}
	\begin{tabular}{cccccccc} %
		\hline
		UGC		& $V_0$ &   $R_\mathrm{PE}$ 	&  $100\alpha$	&$M/L$ 	& $M_b$ & $J_b$ & $j_b$		\\
		\hline	
2023 & 77.2 & 77.2 & 23.2 & 0.78 & 7.7 & 11.6 & 150.8\\
2034 & 29.1 & 12.3 & 6.2 & 1.04 & 13.8 & 20.6 & 149.6\\
2455 & 9.6 & 2.7 & 11.9 & 0.36 & 16.4 & 21.8 & 132.9\\
3371 & 99.1 & 66.0 & -2.8 & 1.87 & 26.0 & 111.0 & 426.2\\
3711 & 92.2 & 12.0 & 0.3 & 0.46 & 7.0 & 14.0 & 198.3\\
3966 & 42.7 & 10.2 & 2.1 & 0.08 & 1.4 & 1.1 & 76.8\\
4173 & 27.9 & 20.3 & 14.0 & 2.88 & 40.1 & 132.0 & 329.0\\
4305 & 43.1 & 64.4 & -4.9 & 0.14 & 7.0 & 7.1 & 100.9\\
4325 & 96.1 & 22.8 & -0.7 & 0.29 & 11.0 & 30.7 & 279.1\\
4499 & 75.8 & 33.4 & -0.2 & 0.35 & 14.2 & 36.3 & 255.4\\
4543 & 56.5 & 2.4 & 0.3 & 0.56 & 96.1 & 708.5 & 736.5\\
5272 & 9.2 & 9.1 & 58.8 & 0.20 & 1.8 & 1.0 & 57.0\\
5414 & 51.6 & 31.4 & 8.8 & 1.11 & 13.3 & 20.2 & 150.7\\
5721 & 90.3 & 27.1 & -1.5 & 0.20 & 5.9 & 13.5 & 227.6\\
5829 & 23.6 & 14.9 & 17.2 & 1.55 & 22.1 & 48.3 & 218.6\\
5918 & 29.9 & 16.0 & 6.6 & 1.22 & 3.1 & 3.8 & 123.9\\
6446 & 68.0 & 18.6 & 2.0 & 1.94 & 42.4 & 125.2 & 295.2\\
6628 & 42.3 & 12.3 & -0.0 & 1.60 & 48.7 & 99.0 & 203.1\\
7047 & 5.9 & 2.3 & 13.8 & 0.25 & 1.2 & 0.5 & 39.8\\
7232 & 11.7 & 10.0 & 46.0 & 0.95 & 1.7 & 0.4 & 25.3\\
7261 & 70.5 & 15.9 & 1.1 & 0.50 & 11.0 & 28.9 & 261.7\\
7323 & 988.5 & 920.0 & -258.1 & 0.54 & 20.1 & 55.0 & 272.5\\
7399 & 89.5 & 20.5 & 1.6 & 0.50 & 7.7 & 18.7 & 241.2\\
7524 & 79.9 & 120.6 & 0.9 & 0.97 & 22.8 & 70.1 & 307.1\\
7559 & 103.6 & 160.3 & -52.4 & 0.25 & 0.7 & 0.2 & 35.5\\
7608 & 68.4 & 50.0 & 4.4 & 0.35 & 6.3 & 12.4 & 195.8\\
7690 & 67.3 & 12.3 & -2.0 & 0.90 & 7.2 & 6.2 & 85.4\\
7866 & 20.1 & 17.9 & 10.9 & 0.40 & 1.4 & 0.6 & 45.6\\
7971 & 571.1 & 535.0 & -281.9 & 1.02 & 7.2 & 5.2 & 72.9\\
8490 & 78.9 & 29.9 & -0.0 & 0.28 & 6.8 & 16.6 & 243.6\\
9211 & 91.0 & 42.4 & -8.0 & 0.52 & 9.8 & 24.5 & 249.3\\
9992 & 30.1 & 6.3 & 1.0 & 0.69 & 3.0 & 2.1 & 69.5\\
10310 & 78.5 & 25.3 & -1.1 & 0.36 & 16.0 & 53.1 & 330.2\\
11861 & 206.9 & 58.2 & -7.7 & 0.86 & 208.8 & 3178.2 & 1522.1\\
12060 & 75.1 & 13.8 & -0.1 & 1.93 & 33.8 & 108.9 & 322.3\\
12632 & 60.4 & 34.6 & 3.4 & 0.19 & 10.6 & 34.0 & 318.7\\
12732 & 59.4 & 20.4 & 5.4 & 1.04 & 40.5 & 273.5 & 673.8\\
       \hline
	\end{tabular}
\end{table*}

\begin{figure}
	\includegraphics[width=\columnwidth]{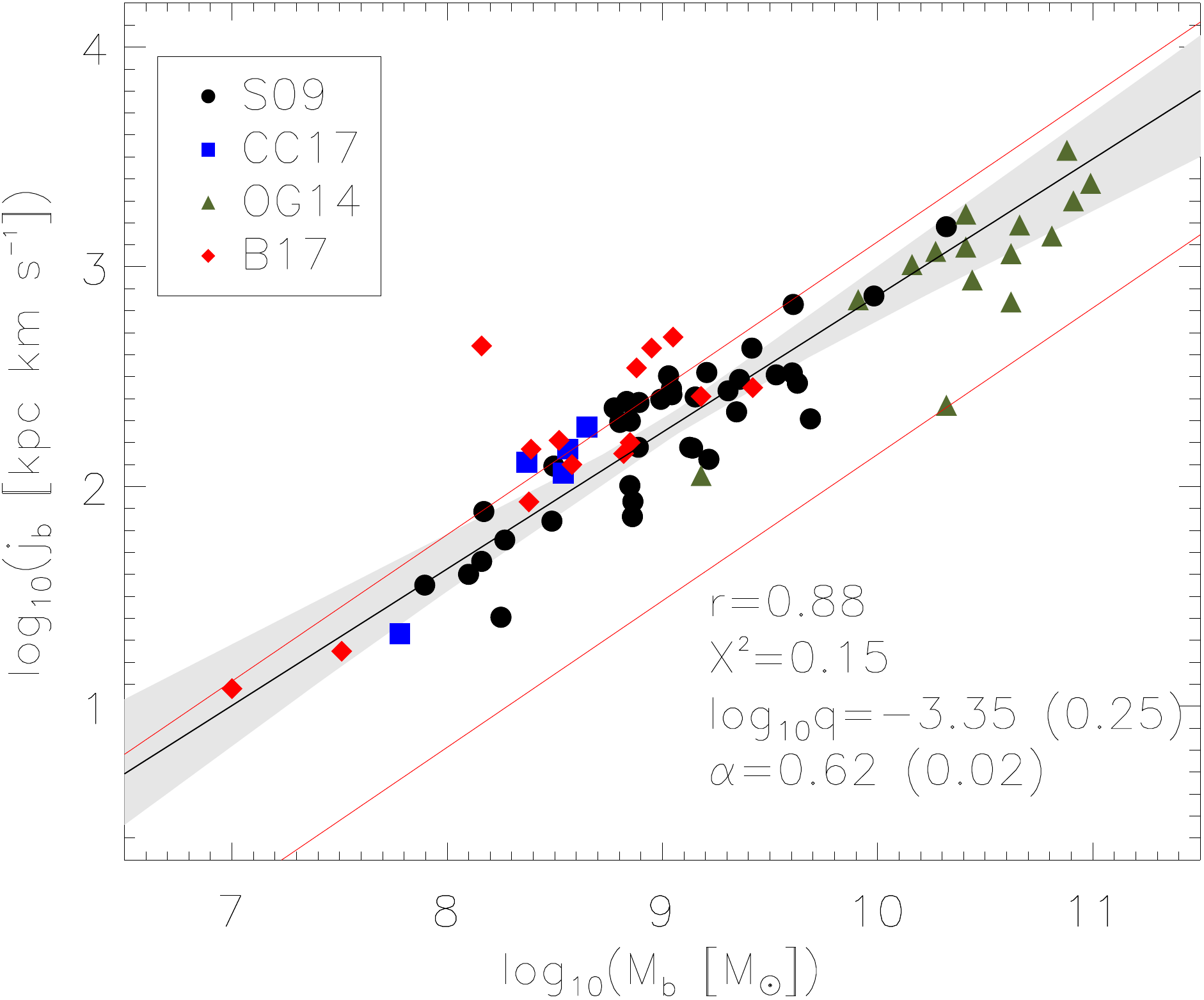}
    \caption{Specific baryon angular momentum  as a function of baryon mass.  The galaxies used in this study are represented by the black circles.  The best-fit relation given by $j_\mathrm{b}=qM_\mathrm{b}^{\alpha}$, with $\alpha=0.62\pm 0.02$ and $\log_{10}q=-3.35\pm 0.25$, is represented by the solid black line.  The Pearson correlation coefficient for the data is $r=0.88$.    The grey-shaded region is based on the jackknifing method used to estimate the uncertainties in $\alpha$ and $\log_{10}q$, it represents the maximum range in \jb\ values from the best-fit relation for any given value of \Mb.  Also shown are the gas-rich galaxies from \citet{CC17} (blue squares), the 16 THINGS spiral galaxies from \citet{OG14} (green triangles) and the 14 LITTLE THINGS galaxies from \citet{Butler_2017} (red diamonds).  The solid red lines represent the theoretically expected range of the \jm\ relation.}
    \label{fig:j-M}
\end{figure}

\begin{figure}
	\includegraphics[width=\columnwidth]{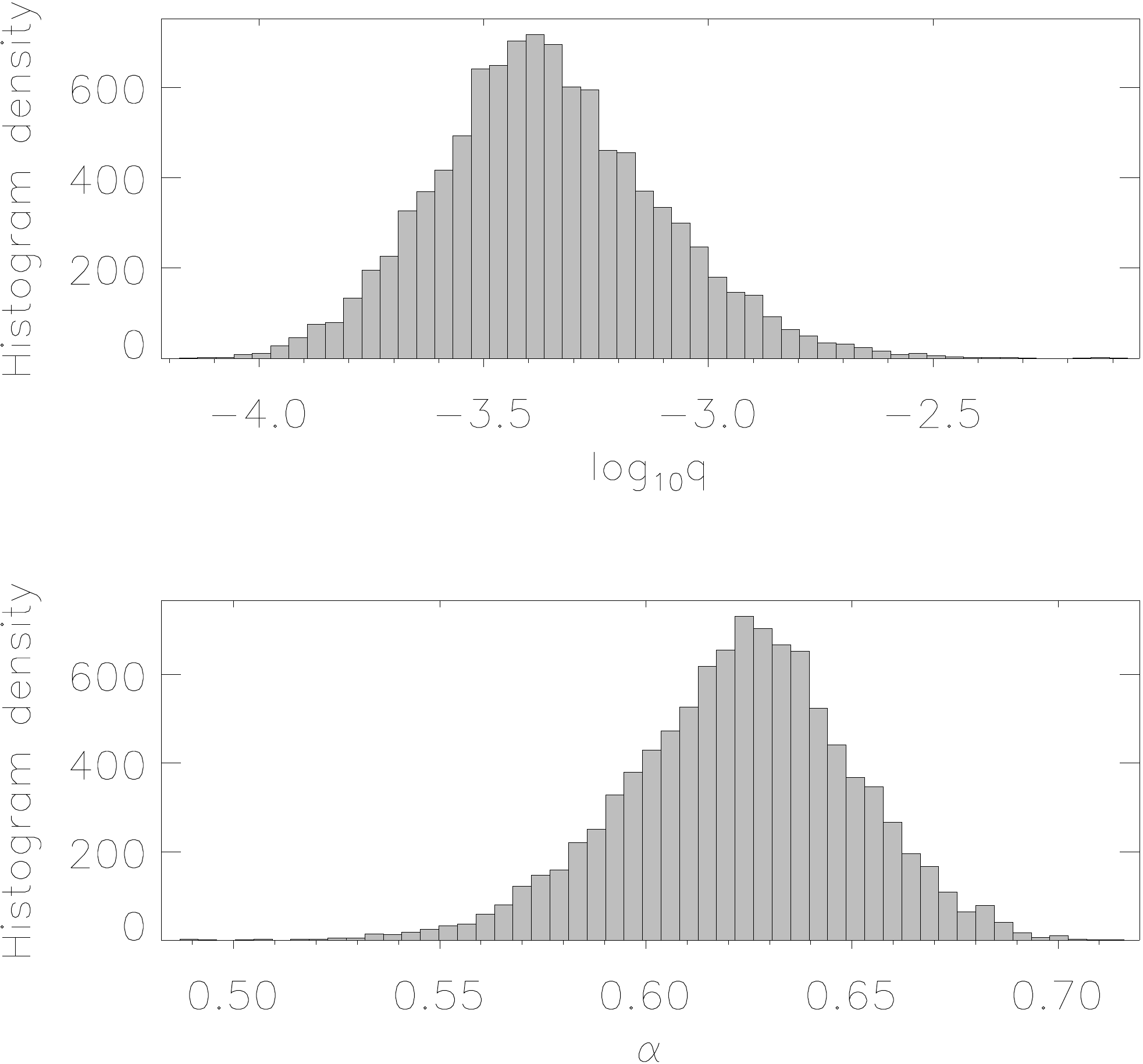}
    \caption{Distribution of best-fit $\alpha$ and $\log_{10}q$  from the jackknifing method used to quantify the measurement uncertainties.}
    \label{fig:parameter_hists}
\end{figure}

The main result from this work, the relation between \jb\ and \Mb,  is shown in Fig.~\ref{fig:j-M}.  Black circles represent the WHISP galaxies from this study, blue squares the gas-rich dwarf galaxies from \citet{CC17}, green triangles the 16 spiral galaxies from THINGS studied by \citet{OG14}, and red diamonds the 14 dwarf Irregular galaxies from the LITTLE THINGS sample \citep{LittleTHINGS} studied by \citet{Butler_2017}.  The jackknifing method was used to derive the linear relation between $\log_{10}M_\mathrm{b}$ and $\log_{10}j_\mathrm{b}$: a first-order polynomial was fit to each of 10$^4$ randomly-selected data subsets of 29 galaxies.  For the relation $j_\mathrm{b}=qM_\mathrm{b}^{\alpha}$, a straight-line in $\log_{10}j_\mathrm{b}$ - $\log_{10}M_\mathrm{b}$ space has slope $\alpha$ and y-axis intercept $\log_{10}q$.  Figure~\ref{fig:parameter_hists} shows the distributions of $\alpha$ and $\log_{10}q$ based on the 10$^4$ jackknife iterations.  The distributions are clearly almost Gaussian in shape.  The mean parameter values are  $\left<\alpha\right>=0.62$ and $\left<\log_{10}q\right>=-3.35$.  These mean values are represented by the solid black line in Fig.~\ref{fig:j-M}.  The standard deviations of the $\alpha$ and $\log_{10}q$ distributions are $\sigma_{\alpha}=0.02$ and $\sigma_{\log_{10}q}=0.25$, respectively.   The correlation between the measured values of \jb\ and those predicted by the best-fit line has a Pearson correlation coefficient of 0.88.  The reduced $\chi^2$ of the fit is 0.15 dex.  For any value of $\log_{10}M_\mathrm{b}$ the grey-shaded region in Fig.~\ref{fig:j-M} represents the maximum range of $\log_{10}j_\mathrm{b}$ values spanned by the 10$^4$  first-order polynomials fitted to the data in the jackknifing process.

\citet{Romanowsky_Fall2012} and \citet{OG14}  both provide equations for the theoretical link between $j$ and $M$.  \citet{OG14} show that for the baryonic case adopting a local $H=70$~\kms~Mpc$^{-1}$, the relation is
\begin{equation}
{j_\mathrm{b}\over 10^3~\mathrm{kpc~km/s}} =1.96~\lambda f_\mathrm{j}f_\mathrm{M}^{-2/3}\left({M_\mathrm{b}\over 10^{10} M_{\odot}}\right)^{2/3},
\end{equation}
where $\lambda$ is the dimensionless spin parameter dealing with the halo angular momentum of a galaxy \citep{Steinmetz_1995}, $f_\mathrm{j}$ is defined as the ratio of the specific angular momentum of the baryons to that of the CDM halo, and $f_\mathrm{M}$ is defined as the ratio of the baryon mass to halo mass.  Using various results from the literature, \citet{OG14} suggest the factor $1.96~\lambda f_\mathrm{j}f_\mathrm{M}^{-2/3}$ can vary between 0.14 and 1.3.  The two red lines in Fig.~\ref{fig:j-M} represent the resulting range in \jb\ values.  Clearly, the $j_\mathrm{b}-M_\mathrm{b}$ relation found in this work is consistent with the theoretical expectation over the entire \Mb\ range probed by the WHISP galaxies.  

The $j_\mathrm{b}-M_\mathrm{b}$ relation found in this work is also consistent with many of the higher mass spiral galaxies studied by \citet{OG14}.  Considering only the power-law slope $\alpha$, it is in fact very similar to the value of $\alpha\approx 2/3$ that \citet{OG14} found for their full sample of THINGS spirals.  \textcolor{black}{However, for a fixed bulge fraction $\beta$, their power-law index is steeper} ($\alpha\approx 1$).  The power-law slope from this work  is also consistent with the results from the \citet{Fall_Romanowsky2013} and \citet{Romanowsky_Fall2012} studies that considered only the stellar component of the specific baryon angular momentum.  \citet{Butler_2017} found their sample of low-mass dwarf Irregular galaxies to deviate from the spiral relation measured by \citet{OG14}.  They show the deviation to be consistent with CDM theory once they account for a decrease in $f_\mathrm{M}$ with decreasing \Mb.  This, they show, has the effect of bending the \jm\ relation at the low \Mb\ end.  The (\jb,  \Mb) measures for the WHISP galaxies presented in this work provide no clear evidence for any such deviation.    

One possible explanation for differences in the results from this work and those of \citet{Butler_2017} may be due to the fact that here the radial profiles used to calculate $j_\mathrm{b}$ are extrapolated only as far as the radius $R_\mathrm{HI}$ at which the face-on \hi\ mass surface density drops to 1~\msun~pc$^{-2}$.  \citet{Butler_2017} produce ``hybrid'' radial profiles for the \hi\ and stellar distributions of their galaxies, as well as their rotation curves, and then extrapolate those profiles out to $15r_\mathrm{HI}$\footnote{$r_\mathrm{HI}$ is the scale length of the exponential fit to the outer part (beyond the central depression) of their measured \hi\ radial profiles.}.  \citet{OG14} explicitly state that limiting $\Sigma_{\mathrm{HI}}$ to $\ge 1$~\msun~pc$^{-2}$ decreases $J_\mathrm{HI}$ and $j_\mathrm{HI}$ by about 20 and 10 per~cent, respectively.  Only one of the galaxies used in this work has its rotation curve in \citet{Swaters2009} measured to a radius beyond $R_\mathrm{HI}$.  Most galaxies have their rotation curves measured out to a radius significantly smaller than $R_\mathrm{HI}$.  Fitting and extrapolating the rotation curves out to $R_\mathrm{HI}$ therefore already constitutes a large extrapolation of the measured data.  It is possible that extrapolating significantly further would lead to large uncertainties in the rotation curves, and hence the corresponding measures of \jb.  UGC~7971 and UGC~7559 are examples of galaxies with measured rotation curves from \citet{Swaters2009} that can easily be flat or rising at radii beyond the last measured point, yet which have their best-fit Polyex models decreasing beyond that point.  

\begin{figure}
	\includegraphics[width=\columnwidth]{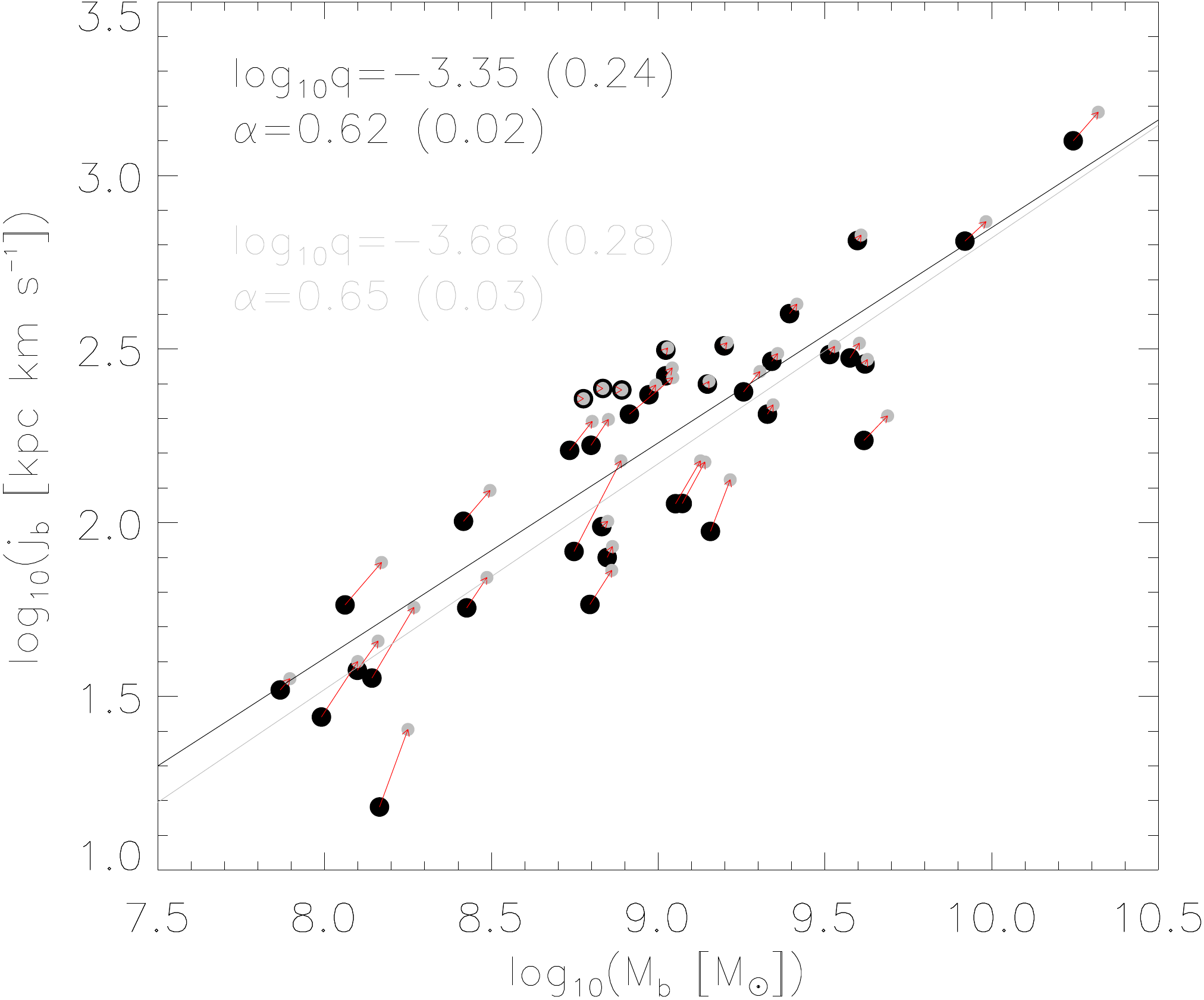}
    \caption{Comparison of measured (\jb, \Mb) pairs for each galaxy used in this study based on $\Sigma_\mathrm{HI}(R)$,  $\Sigma_*(R)$ and $V_\mathrm{PE}(R)$ profiles that are not extrapolated beyond the last measured point on the rotation curve from \citet{Swaters2009} (grey circles) and profiles that are extrapolated to a radius $R_\mathrm{HI}$ (black circles).  Each pair of corresponding points is joined by an arrow.  The black and grey lines represent the best-fit $j_\mathrm{b}-M_\mathrm{b}$ relations for the extrapolated and un-extrapolated profiles, respectively. }
    \label{fig:compare_j-M}
\end{figure}

To provide some handle on the effects of profile extrapolations on the best-fit $j_\mathrm{b}-M_\mathrm{b}$ relation, Fig.~\ref{fig:compare_j-M} shows the change in (\jb, \Mb) measures based on the profiles that are not extrapolated beyond the last measured rotation curve point from \citet{Swaters2009} (grey circles) and profiles that are extrapolated out to $R_\mathrm{HI}$ (black circles, as used in this work) .  Some (\jb, \Mb) pairs do clearly shift fairly significantly within the $\log_{10}M_\mathrm{b}$ - $\log_{10}j_\mathrm{b}$ plane, especially for $M_\mathrm{b}\lesssim 10^{9.25}$~\msun.  However, most shifts are roughly parallel to the best-fit relation, which therefore hardly changes.  Using the un-extrapolated profiles, the best-fit relation yields $\left<\alpha\right>=0.65 \pm 0.03$ and $\left<\log_{\mathrm{10}}q\right>=-3.68\pm 0.28$, with a Pearson correlation coefficient of 0.87 and a reduced $\chi^2$ of 0.17 dex.  This relation has larger parameter uncertainties and more scatter.  This could be due to the fact that the \jb$(R)$ measures have not sufficiently converged at  the last-measured points on the original (un-extrapolated) rotation curves from \citet{Swaters2009}.  It would certainly be much more preferable to directly measure the various constituent profiles required to calculate \jb\ and \Mb\ out to radii that ensure both quantities converge to constant values.  Future projects such as the MHONGOOSE survey (\url{http://mhongoose.astron.nl/}) to be carried out on the new MeerKAT telescope might provide sufficiently sensitive \hi\ observations of nearby galaxies to carry out a highly reliable study of their $j_\mathrm{b}-M_\mathrm{b}$ properties.

\section{Low-resolution systematics}\label{systematics}
\begin{figure}
	\includegraphics[width=\columnwidth]{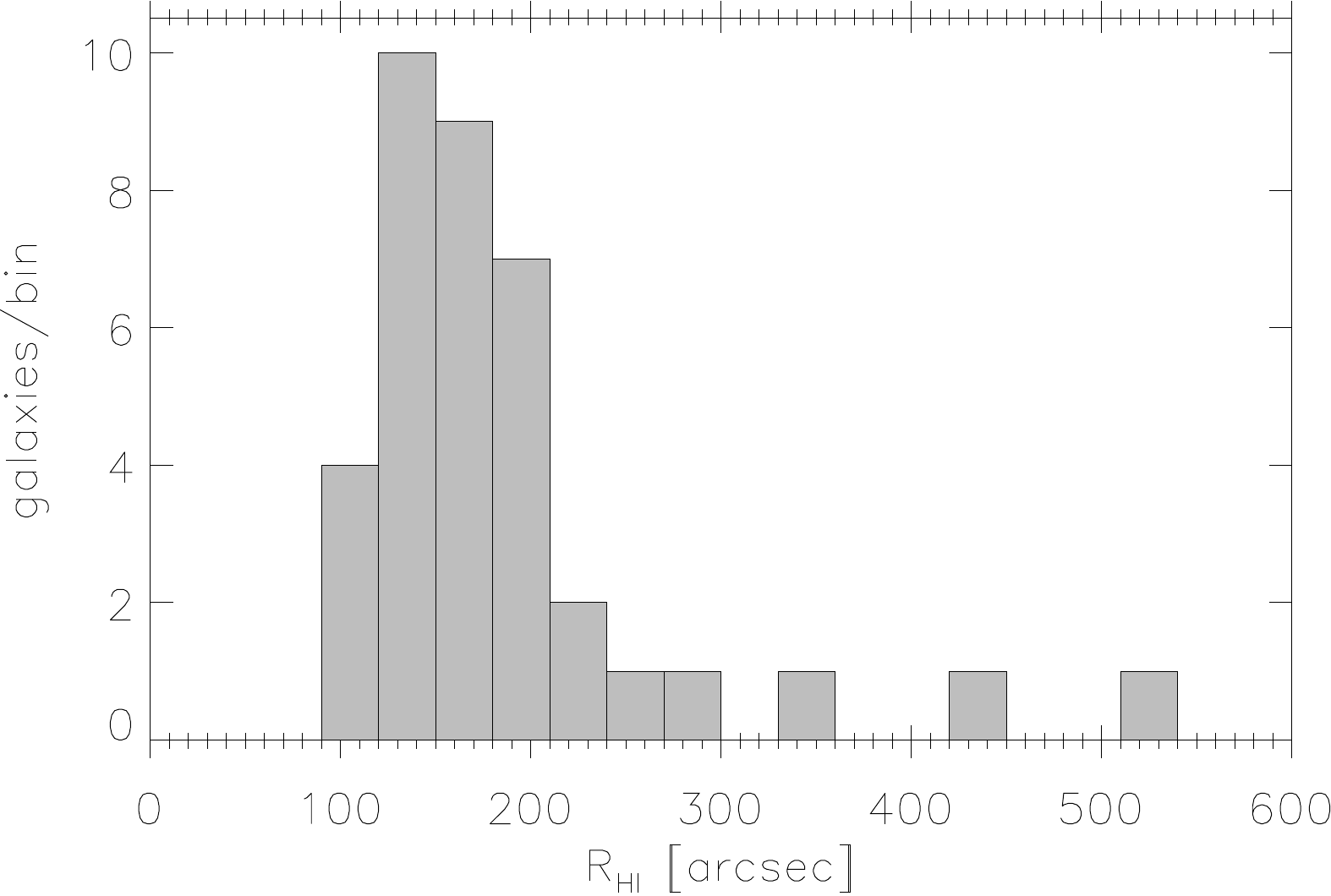}
    \caption{Distribution of HI radii for the galaxies used in this study.  Given the 30~arcsec spatial resolution of the \hi\ imaging, the majority of galaxies have 4 - 7 \hi\ beams across their \textit{semi}-major axis. }
    \label{fig:RHI}
\end{figure}

The main difference between the WHISP data used in this work and THINGS data used by other investigators is the spatial resolution.  The JVLA \hi\ data cubes used by \citet{OG14} and \citet{Butler_2017} have spatial resolutions of $\sim 6''$, resulting in the galaxies being resolved by many beams ($\gtrsim 200$) across their major axes.  In contrast, the WHISP \hi\ total intensity maps used in this work have a resolution of 30~arcsec.  Figure.~\ref{fig:RHI} shows the distribution of their \hi\ radii $R_\mathrm{HI}$.  The large majority of galaxies have 4 - 7 \hi\ beams across their \textit{semi}-major axis.  This is as much as an order of magnitude (or more) fewer than the typical number of beams across the THINGS galaxies.  A particular concern, therefore, is the possibility of the \jb\ and \Mb\ measures presented in this study being systematically affected by the low spatial resolution.  

\begin{figure*}
	\includegraphics[width=1.8\columnwidth]{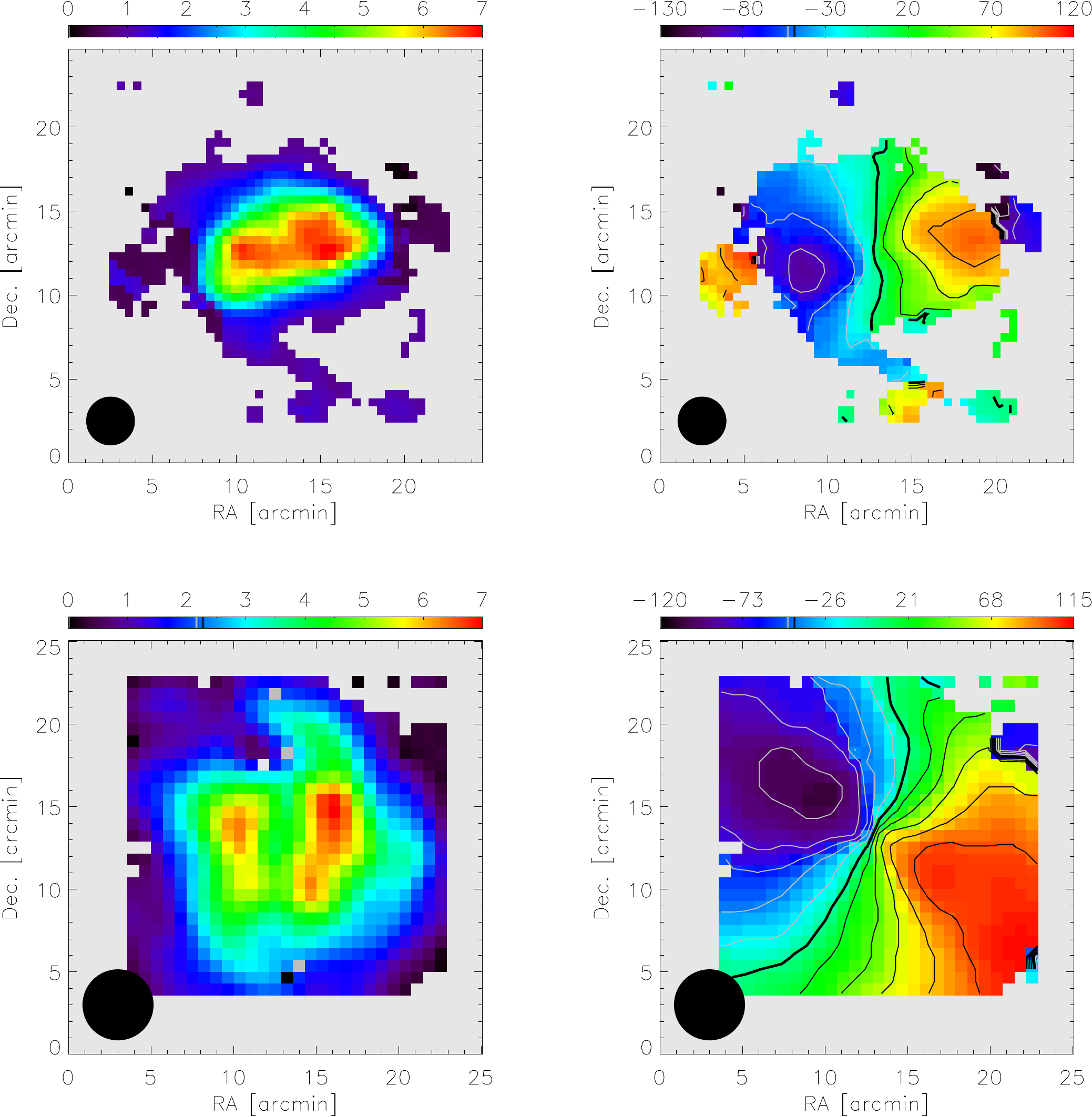}
    \caption{\hi\ total intensity maps (left) and \hi\ velocity fields (right) for NGC~925 (top) and NGC~6946 (bottom) based on the spatially-smoothed versions of the THINGS \hi\ data cubes discussed in Sec.~\ref{systematics}.  These maps are used to derive the radial profiles shown in Fig.~\ref{fig:N925_6946_profiles}.  The colour bar above each map gives the intensity scale in units of \msun~pc$^{-2}$ and \kms\ for the total intensity and velocity maps, respectively.  In each panel, the black-filled circle in the bottom left represents the half-power size of the Gaussian PSF.  For the velocity fields, the thick black contour at the centre is at 0~\kms.  The thin black/grey contours are spaced by 30/20~\kms\ for NGC~925/6946.}
    \label{fig:smoothed_maps}
\end{figure*}

In this section, two THINGS galaxies from the \citet{OG14} study are used to quantify the combined systematic effects of low spatial resolution on the \jb\ and \Mb\ measures.  NGC~925 is chosen as a low-mass system with M$_\mathrm{HI}=45.5\times 10^8$~\msun\ \citep{THINGS_Walter} and  M$_*= 1.02\times 10^{10}$~\msun\ \citep{THINGS_deBlok} that also has a low maximum rotation speed of $\sim 119$~\kms\ at $R\sim 13$~kpc \citep{THINGS_deBlok} that is similar to the typical galaxy used in this work.  NGC~6964 is chosen as a higher-mass system with M$_\mathrm{HI}=41.5\times 10^8$~\msun\ \citep{THINGS_Walter} and  M$_*= 5.88\times 10^{10}$~\msun\ \citep{THINGS_deBlok} with a higher maximum rotation speed of $\sim 200$~\kms\ for $5~\mathrm{kpc} \lesssim R \lesssim 17~\mathrm{kpc}$ \citep{THINGS_deBlok}.  The naturally-weighted \hi\ data cube of each galaxy was downloaded from the THINGS public data repository (\url{http://www.mpia.de/THINGS/Data.html}).  Each cube was spatially smoothed on a channel-by-channel basis to a resolution corresponding to $\sim 5$ beams spanning the semi-major axis of the galaxy.  The \hi\ mass-size relation from \citet{Obresch_2009} was used together with the total \hi\ masses of the galaxies from \citet{THINGS_Walter} to calculate $R_\mathrm{HI} = 437$~arcsec and 646~arcsec for NGC 925 and 6946, respectively.  NGC~925 was therefore smoothed to 87~arcsec spatial resolution and NGC~6946 to 129~arcsec.  The smoothed cubes were spatially re-gridded to have pixel sizes equal to a third of the spatial resolution.  For each smoothed, re-gridded cube, a third-order Gauss-Hermite polynomial was fit to all of its line profiles.  All fitted profiles with peak amplitude greater than three times the RMS of the emission in a line-free channel  were used to generate an \hi\ total intensity map by spectrally integrating the profiles, and an \hi\ velocity field by selecting from each profile the line-of-sight velocity corresponding to the peak amplitude.  Figure ~\ref{fig:smoothed_maps} shows the \hi\ maps for NGC~925 and NGC~6946.  These \hi\ maps are used to calculate values of \jb\ and \Mb\ for the two galaxies that can be compared to the corresponding values from \citet{OG14} based on the full-resolution THINGS imaging.  

\begin{figure}
	\includegraphics[width=\columnwidth]{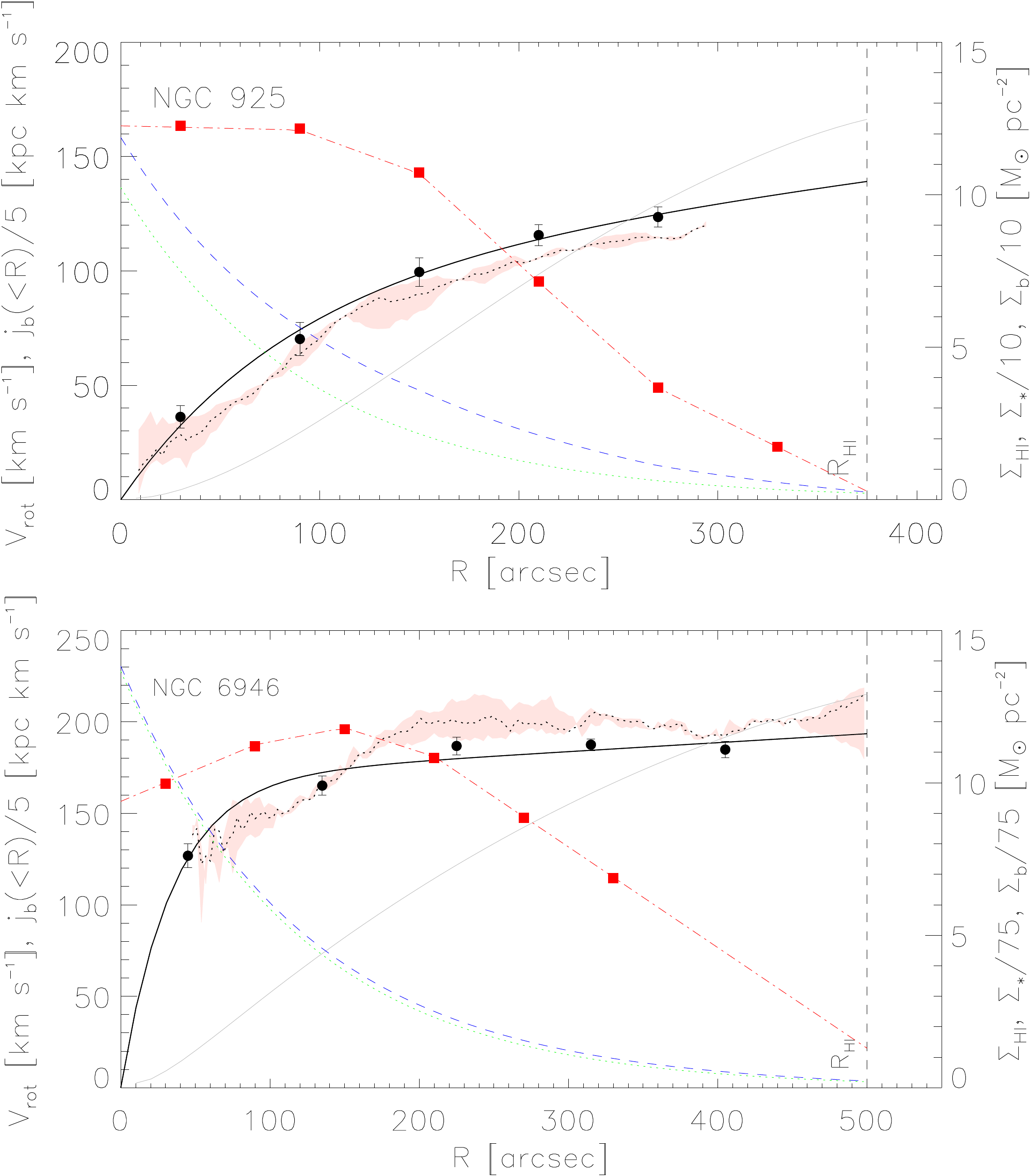}
    \caption{Various radial profiles for NGC~925 (top) and NGC~6946 (bottom) derived from the \hi\ maps shown in Fig.~\ref{fig:smoothed_maps}.  Black-filled circles represent the rotation curves from the tilted ring models fit to the velocity fields.  The solid black curve is the corresponding best-fit Polyex model (Eqn.~\ref{eqn:polyex}).  For comparison, the rotation curves derived by \citet{THINGS_deBlok} using the full-resolution THINGS maps are shown as black dotted curves.  The salmon-shaded regions represent the difference in rotational velocities for the approaching and receding halves of the galaxies.  The red-filled squares joined by the red dot-dash lines represent the \hi\ surface densities $\Sigma_\mathrm{HI}(R)$ derived from the smoothed \hi\ total intensity maps. The green-dotted curves represent the exponential component of the stellar disk, the blue-dashed curves the baryon mass radial profile $\Sigma_\mathrm{b}(R)$.  The solid grey curves represent the cumulative specific baryon angular momentum $j_\mathrm{b}(\le R)$.  The vertical black-dashed lines represent the \hi\ radius $R_\mathrm{HI}$ based on the total \hi\ mass of the galaxy.}
    \label{fig:N925_6946_profiles}
\end{figure}

An \hi\ mass surface density profile was generated for each galaxy using exactly the same method described in Sec.~\ref{HI_surface_density_profiles} for the WHISP galaxies.  The $\Sigma_\mathrm{HI}(R)$ profiles are shown as the red-filled circles joined by dot-dash lines in Fig.~\ref{fig:N925_6946_profiles}.  The exponential stellar disks of the galaxies were modelled by fitting exponential functions to the outer parts (excluding the central bulge) of their measured stellar mass surface density profiles, kindly provided by the THINGS collaboration.  The $\Sigma_\mathrm{*}(R)$ profiles are shown as the green-dotted curves in Fig.~\ref{fig:N925_6946_profiles}.  The baryon mass surface density profiles $\Sigma_\mathrm{b}(R)$ calculated according to Eqn~\ref{eqn_Mb} are shown as the blue-dashed curves in Fig.~\ref{fig:N925_6946_profiles}.  

The {\sc rotcur} routine in {\sc gipsy} \citep{GIPSY} was used to fit a tilted ring model to each of the velocity fields shown in Fig.~\ref{fig:smoothed_maps} in order to generate rotation curves.  The same, simple approach was used for each galaxy: First, an unconstrained model in which only the expansion velocity was fixed (to 0~\kms) was used  to generate a feel for the natural radial variations of the various parameters.  The mean values of the dynamical centre, systemic velocity, position angle and inclination were determined and used as fixed values in the second {\sc rotcur} iteration which then yielded the final rotation curve.  The fixed values of inclination and position angle used for NGC~925 were 65.9\deg\ and 288.5\deg, respectively.  These values are very close to the distribution of inclination and position angle that were ultimately adopted by \citet{THINGS_deBlok} to derive their final rotation curve (see their Fig.~67).  The fixed inclination for NGC~6946 was 35.0\deg\ and the position angle 242.6\deg\ - again very similar to the distributions used by \citet{THINGS_deBlok} (see their Fig.~83).  Figure~\ref{fig:N925_6946_profiles} shows as black-filled circles the final NGC~925 and NGC~6946 rotation curves from the tilted ring models.  The best-fitting Polyex model for each is shown as the solid black curve.  For comparison, the THINGS rotation curves from \citet{THINGS_deBlok} are shown as the black dotted curves.  The effects of the low spatial resolution of the data are arguably quite clearly evident.  The derived rotation velocities for NGC~925 are larger than the THINGS rotation velocities by $\sim 10$~\kms\ at nearly all radii.  For NGC~6946, the derived rotation velocities are lower than the THINGS rotation curve over most of the galaxy ($R\ge 225$~arcsec).

The extrapolated Polyex model rotation curves and baryon mass surface density profiles  based on the low-resolution THINGS imaging (i.e., those shown in Fig.~\ref{fig:N925_6946_profiles}) are used in equations \ref{eqn:Mb} and \ref{eqn:J} to calculate \jb\ and \Mb\ measures for NGC~925 and NGC~6946.  Table~\ref{tab:comparison} shows the newly calculated \jb\ and \Mb\ measures in the left column and the corresponding values from \citet{OG14} in the right column.  Despite having significantly lower spatial resolution, the modified THINGS maps yield very similar measures to those obtained by \citet{OG14}.  When considering the logarithms of the parameters, the largest percentage difference is 2.9~per~cent.  Given that the \citet{Swaters2009} rotation curves are certainly of much higher quality than those derived here for NGC~926 and NGC~6946, the results from this section  strongly suggest the \jb\ and \Mb\ measures derived in this work contain no significant systematic errors linked to the relatively low-resolution (compared to THINGS) of the WHISP imaging.  The best-fit results of $\alpha=0.62\pm0.02$ and $\log_{10}q=-3.35\pm 0.25$  presented in Sec.~\ref{results} are almost certainly highly reliable and accurate. 

\begin{table}
\centering
\caption{\jb\ and \Mb\ measures for NGC~925 and NGC~6946.  Column 2 gives the measures based on the smoothed THINGS \hi\ maps discussed and presented  in Sec.~\ref{systematics}.  Column 3 gives the corresponding values from \citet{OG14} based on the full-resolution THINGS imaging.  Column 4 gives the absolute differences relative to the \citet{OG14} measures.}
	
\label{tab:comparison}
	\begin{tabular}{cccc} %
		\hline
		                                     	& This work  	&OG14	&$\%$ diff\\
		\hline	
		NGC~925				&			&		&		\\
		$\log_{10}j_\mathrm{b}$ 	&2.92 		&3.01 	&2.9		\\
		$\log_{10}M_\mathrm{b}$	&10.23		&10.16	&0.7		\\
		\hline	
		NGC~6946			&			&		&		\\
		$\log_{10}j_\mathrm{b}$ 	&3.03 		&3.06 	&1.0		\\
		$\log_{10}M_\mathrm{b}$	&10.86		&10.62	&2.3		\\
       \hline
	\end{tabular}
\end{table}


\section{Summary and Conclusions}\label{conclusions}
This work uses a sample of 37 galaxies from the WHISP survey with spatially and kinematically resolved \hi\ kinematics to study the relation between  specific baryon angular momentum \jb\ and total baryon mass \Mb\ over the range $10^8\lesssim M_\mathrm{b}/M_{\odot}\lesssim 10.5$.  This study roughly doubles the number of galaxies with $M_\mathrm{b}\lesssim 10^{10}$~\msun\ used to study the $j_\mathrm{b}-M_\mathrm{b}$ relation.  

The best-fit relation is given by $j_\mathrm{b}=qM_\mathrm{b}^{\alpha}$, with $\alpha=0.62\pm 0.02$ and $\log_{10}q=-3.35\pm 0.25$.  These best-fit parameters are shown to be robust against the effects of extrapolating the constituent radial profiles used to calculate \jb\ and \Mb.  They are also shown to be unaffected by the spatial resolution of the WHISP \hi\ imaging.  The best-fit \jm\ relation is consistent with the theoretically expected $j_\mathrm{b}\propto M_\mathrm{b}^{2/3}$ relation, at least over the \Mb\ range spanned by the WHISP galaxies.  It is also consistent with the (\jb, \Mb) measures of most of the  higher mass spiral galaxies from \citet{OG14}.  Unlike the results for a sample of dwarf-Irregular galaxies from \citet{Butler_2017}, the results presented here do not provide an clear evidence for \jb\ measures that are  systematically higher than the expected theoretical relation or those expected from the $j_\mathrm{b}-M_\mathrm{b}$ scaling of spiral galaxies.   

\section{Acknowledgements}
The anonymous referee is thanked for providing comments and feedback that improved the overall quality of the paper.  EE acknowledges the support of the South African National Astrophysics and Space Science Programme (NASSP).


\bibliographystyle{mnras}


\appendix

\section{Radial profiles}\label{appendixA}
The next pages show the various radial profiles generated and used in this work.  Each panel corresponds to a different galaxy.  For each galaxy the details are as follows: the \hi\ mass surface density profile $\Sigma_\mathrm{HI}(R)$ measured from the 30~arcsec WHISP \hi\ total intensity map is represented by the red dash-dot curve, the measured rotation curve from \citet{Swaters2009} is represented by the black circles, the model rotation curve is represented by the solid black curve, the surface brightness profile for the exponential disk component of the stellar disk $\Sigma_*(R)$ is represented as the green dotted curve, the total baryonic mass surface density profile $\Sigma_\mathrm{b}(R)$ is represented by the blue dashed curve, the cumulative baryonic specific angular momentum profile $j_\mathrm{b}(<R)$ is represented by the grey solid curve.  

The UGC number of the galaxy is given above each panel.  Inside each panel is given: 1) the physical length scale in units of kpc corresponding to the 30~arcsec resolution of the \hi\ imaging, 2) its rotation curve quality parameter as detailed in \citet{Swaters2009}, 3) its specific angular momentum in units of kpc~\kms\ at radius $R_\mathrm{HI}$, and 4) $R_\mathrm{HI}$ in units of kpc.

\begin{figure*}
	\includegraphics[width=1.8\columnwidth]{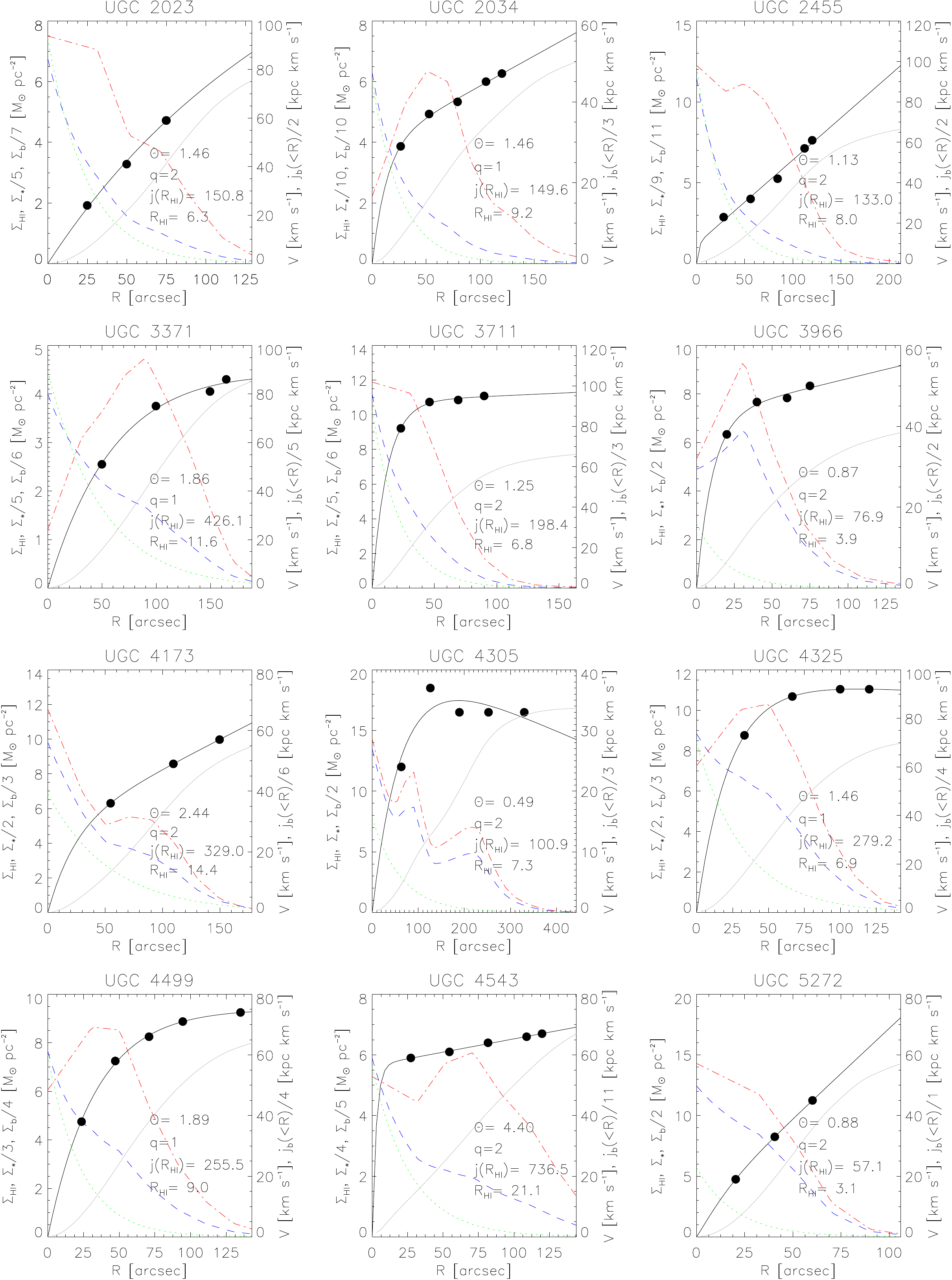}
    \caption{Various radial profiles for each galaxy used in this study.   The radial profiles for $\Sigma_{HI}$, $\Sigma_{*}$, $\Sigma_{b}$, $j_\mathrm{b}(<R)$ and $V$ are represented by the red, green, blue, grey and black curves, respectively.  See the start of Appendix~\ref{appendixA} for a detailed description the content in each panel.}
    \label{fig:profiles1}
\end{figure*}

\begin{figure*}
	\includegraphics[width=1.8\columnwidth]{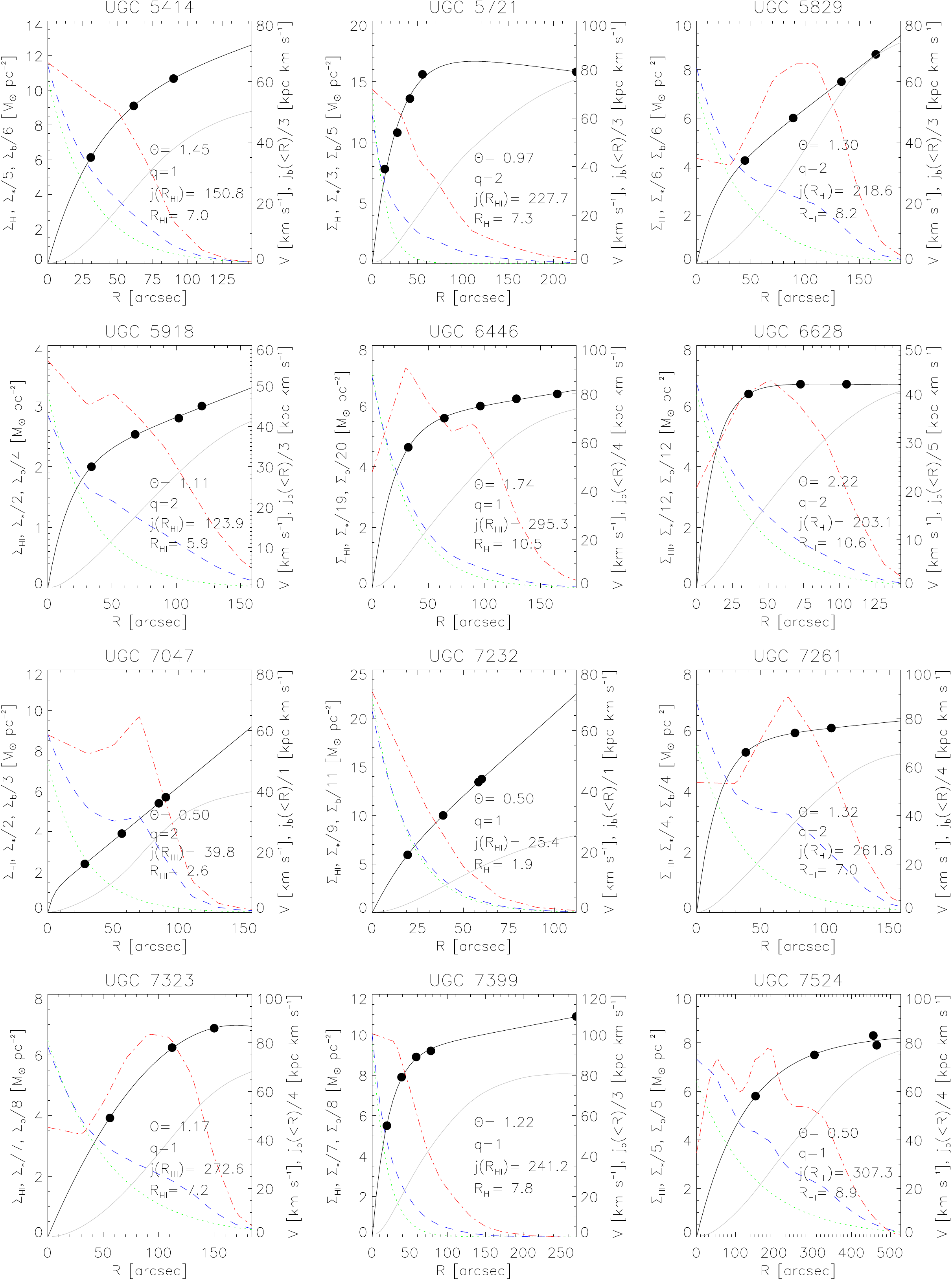}
    \caption{Various radial profiles for the galaxies used in this study.  See Fig.~\ref{fig:profiles1} caption and the start of Appendix~\ref{appendixA} for more details.}
    \label{fig:profiles2}
\end{figure*}

\begin{figure*}
	\includegraphics[width=1.8\columnwidth]{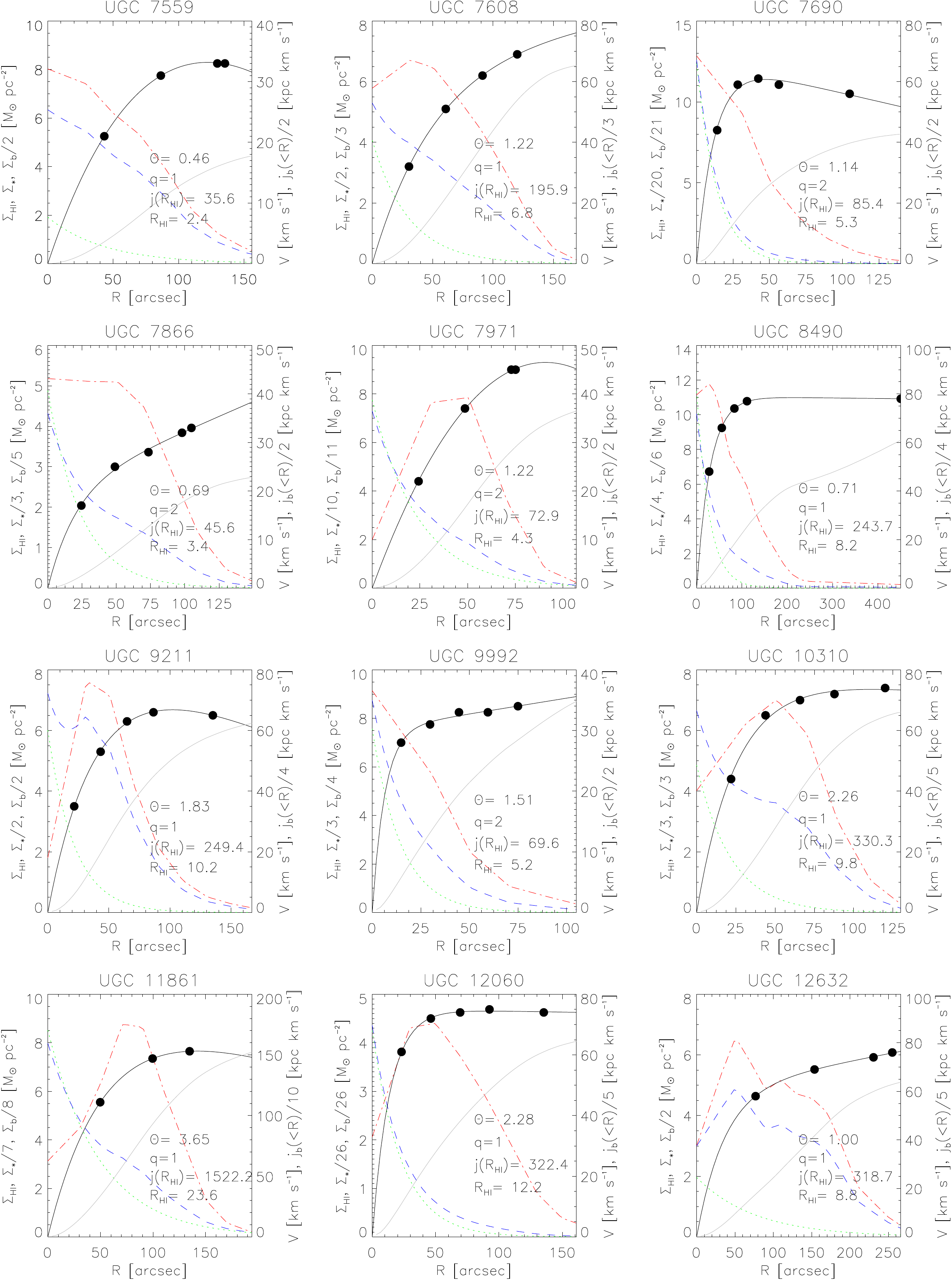}
    \caption{Various radial profiles for the galaxies used in this study.  See Fig.~\ref{fig:profiles1} caption and the start of Appendix~\ref{appendixA} for more details.}
    \label{fig:profiles3}
\end{figure*}

\begin{figure}
	\includegraphics[width=0.55\columnwidth]{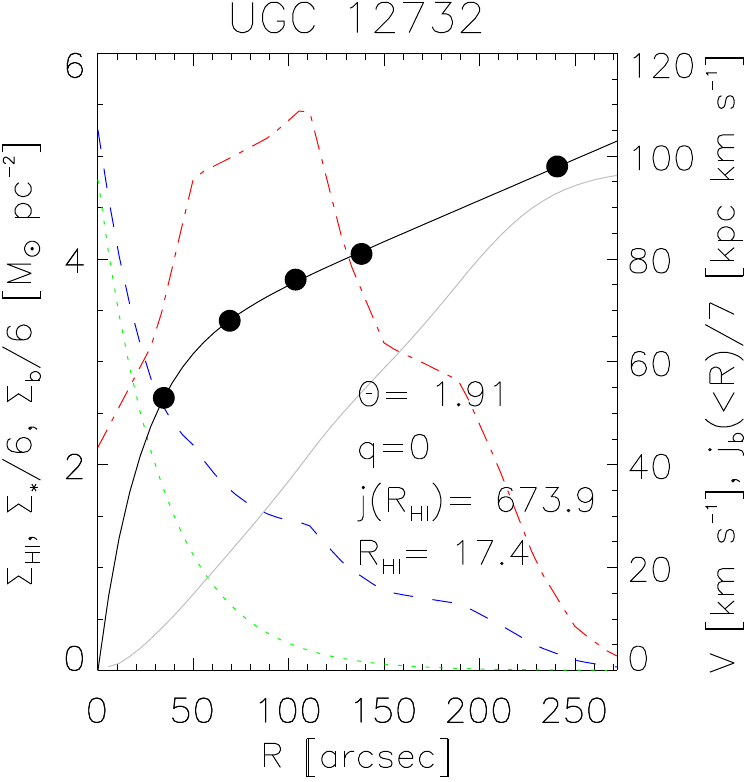}
    \caption{Various radial profiles for UGC 12732 used in this study.  See Fig.~\ref{fig:profiles1} caption and the start of Appendix~\ref{appendixA} for more details.}
    \label{fig:profiles4}
\end{figure}


\bsp	
\label{lastpage}
\end{document}